\documentclass[prx,aps,amssymb,twocolumn,reprint,superscriptaddress]{revtex4-1}
\usepackage{amsmath}
\usepackage{amssymb}
\usepackage{amsthm}
\usepackage{amsfonts}
\usepackage{listings}
\usepackage{enumerate}
\usepackage{latexsym}
\usepackage{color}
\usepackage{xcolor}
\usepackage{bm}
\usepackage{hyperref}
\hypersetup{
 pdfnewwindow=true, colorlinks=true,
 linkcolor=blue, anchorcolor=blue,
 citecolor=blue, filecolor=blue,
 menucolor=blue, urlcolor=blue}

\usepackage{psfrag}

\usepackage{graphicx}
\usepackage{subfigure}
\usepackage{lipsum}
\usepackage{float}
\usepackage{array}
\usepackage{soul} 
\usepackage{titlesec}

\newcommand{\sect}[1]{{\noindent\large\bf {#1}}\par}
\newcommand{\subsect}[1]{{\noindent\normalsize\bf {#1}}\par}

\newcommand{\beginsupplement}{%
        \setcounter{table}{0}
        \renewcommand{\thetable}{S\arabic{table}}%
        \setcounter{figure}{0}
        \renewcommand{\thefigure}{S\arabic{figure}}%
        \setcounter{equation}{0}
        \renewcommand{\theequation}{S\arabic{equation}}
     }

\begin{document}
\soulregister{\cite}7
\soulregister{\ref}7
\tolerance 10000

\draft
\title{Materials discovery acceleration by using condition generative methodology}

\author{Caiyuan Ye}
\affiliation{
 Beijing National Laboratory for Condensed Matter Physics and Institute of Physics, Chinese Academy of Sciences, Beijing 100190, China\\
}%
\affiliation{University of Chinese Academy of Sciences, Beijing 100049, China \\}

\author{Yuzhi Wang}
\affiliation{
 Beijing National Laboratory for Condensed Matter Physics and Institute of Physics, Chinese Academy of Sciences, Beijing 100190, China\\
}%
\affiliation{University of Chinese Academy of Sciences, Beijing 100049, China \\}

\author{Xintian Xie}
\affiliation{
 Shanghai Key Laboratory of Molecular Catalysis and Innovative Materials, Department of Chemistry, Key Laboratory of Computational Physical Science (Ministry of Education), Fudan University, Shanghai 200433, China\\}

\author{Tiannian Zhu}
\affiliation{
 Beijing National Laboratory for Condensed Matter Physics and Institute of Physics, Chinese Academy of Sciences, Beijing 100190, China\\
}%
\affiliation{University of Chinese Academy of Sciences, Beijing 100049, China \\}

\author{Jiaxuan Liu}
\affiliation{
 Beijing National Laboratory for Condensed Matter Physics and Institute of Physics, Chinese Academy of Sciences, Beijing 100190, China\\
}%
\affiliation{University of Chinese Academy of Sciences, Beijing 100049, China \\}

\author{Yuqing He}
\affiliation{
 Beijing National Laboratory for Condensed Matter Physics and Institute of Physics, Chinese Academy of Sciences, Beijing 100190, China\\
}%

\author{Lili Zhang}
\affiliation{
 Chongqing Bitmap Information Technology Co. Ltd, Chongqing 402760, China\\
}%

\author{Junwei Zhang}
\affiliation{
 Chongqing Bitmap Information Technology Co. Ltd, Chongqing 402760, China\\
}%

\author{Zhong Fang}
\affiliation{
 Beijing National Laboratory for Condensed Matter Physics and Institute of Physics, Chinese Academy of Sciences, Beijing 100190, China\\
}%
\affiliation{University of Chinese Academy of Sciences, Beijing 100049, China \\}

\author{Lei Wang}
\affiliation{
 Beijing National Laboratory for Condensed Matter Physics and Institute of Physics, Chinese Academy of Sciences, Beijing 100190, China\\
}%

\author{Zhipan Liu}
 \email{zpliu@fudan.edu.cn}
\affiliation{
 Shanghai Key Laboratory of Molecular Catalysis and Innovative Materials, Department of Chemistry, Key Laboratory of Computational Physical Science (Ministry of Education), Fudan University, Shanghai 200433, China\\}

\author{Hongming Weng}
 \email{hmweng@iphy.ac.cn}
\affiliation{
 Beijing National Laboratory for Condensed Matter Physics and Institute of Physics, Chinese Academy of Sciences, Beijing 100190, China\\
}%
\affiliation{University of Chinese Academy of Sciences, Beijing 100049, China \\}

\author{Quansheng Wu}
 \email{quansheng.wu@iphy.ac.cn}
\affiliation{
 Beijing National Laboratory for Condensed Matter Physics and Institute of Physics, Chinese Academy of Sciences, Beijing 100190, China\\
}%
\affiliation{University of Chinese Academy of Sciences, Beijing 100049, China \\}

\begin{abstract}
With the rapid advancement of AI technologies, generative models have been increasingly employed in the exploration of novel materials. By integrating traditional computational approaches such as density functional theory (DFT) and molecular dynamics (MD), existing generative models—including diffusion models and autoregressive models—have demonstrated remarkable potential in the discovery of novel materials. However, their efficiency in goal-directed materials design remains suboptimal. In this work we developed a highly transferable, efficient and robust conditional generation framework, PODGen, by integrating a general generative model with multiple property prediction models. Based on PODGen, we designed a workflow for the high-throughput crystals conditional generation which is used to search new topological insulators (TIs). Our results show that the success rate of generating TIs using our framework is 5.3 times higher than that of the unconstrained approach. More importantly, while general methods rarely produce gapped TIs, our framework succeeds consistently—highlighting an effectively $\infty$ improvement. This demonstrates that conditional generation significantly enhances the efficiency of targeted material discovery. Using this method, we generated tens of thousands of new topological materials and conducted further first-principles calculations on those with promising application potential. Furthermore, we identified promising, synthesizable topological (crystalline) insulators such as $\mathrm{CsHgSb}$, $\mathrm{NaLaB_{12}}$, $\mathrm{Bi_4Sb_2Se_3}$, $\mathrm{Be_3Ta_2Si}$ and $\mathrm{Be_2W}$.
\end{abstract}

\maketitle

\section{Introduction}

New materials play a crucial role in industrial and technological fields\cite{zhang2021dealing}, offering unique properties that drive more efficient and sustainable solutions\cite{hu2021research}. Crystalline materials, with their highly ordered structures, excel in areas such as electronics, optoelectronics, and medicine, providing significant support for technological advancements\cite{disa2021engineering,butt2021recent,chaudhary2022advances}. However, traditional experimental and theoretical computation methods are increasingly unable to meet growing demands\cite{agrawal2016perspective}.

With the rapid advancement of AI technology over the past decade, new research paradigms have been introduced into the discovery of novel crystal materials, offering the potential to overcome the limitations of traditional methods\cite{paradigms,review}. Well-developed predictive machine learning models have already demonstrated their ability to facilitate the rapid and accurate screening of crystal structures\cite{cgcnn,megnet,alignn,dimenet,gemnet,coNGN,liang2023material}, assisting, accelerating, and even replacing first-principles calculations\cite{m3gnet,chgnet,gnome,liang2024cluster,yang2024mattersim,omat24,AlphaNet,hamgnn,hamgnn2,deeph2,deephe3,dpmoire}. In recent years, various generative machine learning models have been applied to the exploration of new crystal structures. For example, diffusion-based models\cite{cdvae,diffcsp,diffcsp++,joshi2025all} such as CDVAE, autoregressive models\cite{schnet,cifllm,crystalformer} such as CrystalFormer, flow-based models\cite{miller2024flowmm,luo2024crystalflow} such as FlowMM, as well as several other models\cite{qiu2024vqcrystal,sriram2024flowllm}. These general generative machine learning models primarily focus on learning the distribution of crystal structures from training datasets, enabling the sampling of novel structures. Alongside these general generative models, conditional generative models have been developed to generate crystal structures tailored to specific target properties. Examples include MatterGen\cite{zeni2023mattergen}, MatterGPT\cite{chen2024mattergpt}, Con-CDVAE\cite{con-cdvae}, and Cond-CDVAE\cite{cond-cdvae}. It is also noteworthy that recent efforts have employed reinforcement learning\cite{CFRL,RLTI} or active learning\cite{active1,active2} to achieve conditional generation of crystal structures.

While some studies have explored the application of generative models in crystal structure discovery\cite{cdvae_super,cdvae_2d,cdvae_topo,wyck_gen}, materials with desirable physical properties and practical applications often constitute only a small fraction of known structures. In such cases, conditional generative models offer a more efficient approach than general generative models by guiding the search toward structures that meet specific criteria.

In this paper, we propose a conditional generation framework named PODGen, which means using \textbf{P}redictive models to \textbf{O}ptimize the \textbf{D}istribution of the \textbf{Gen}erative model for conditional generation. It can be applied to various generative and predictive models, effectively improving the success rate of generation. Additionally, we have designed a workflow for high-throughput generation of crystal structures, including structure optimization, property verification, and structure deduplication. We demonstrate its application in generating topological insulators, which are crystalline materials with special electronic band structures that enable the formation of protected surface states, exhibiting unique electrical and spin-related properties\cite{tqcwang}. And 19324 topological insulators and topological crystalline insulators have been generated, with further first-principles calculations performed on promising materials with potential practical applications, which found 12 new dynamically stable (no imaginary phonon modes) crystal structures with desirable properties, among which 5 are located at the bottom of the potential energy surface (PES).

\section{Method}
\begin{figure*}[t]
\includegraphics[width=0.9\textwidth]{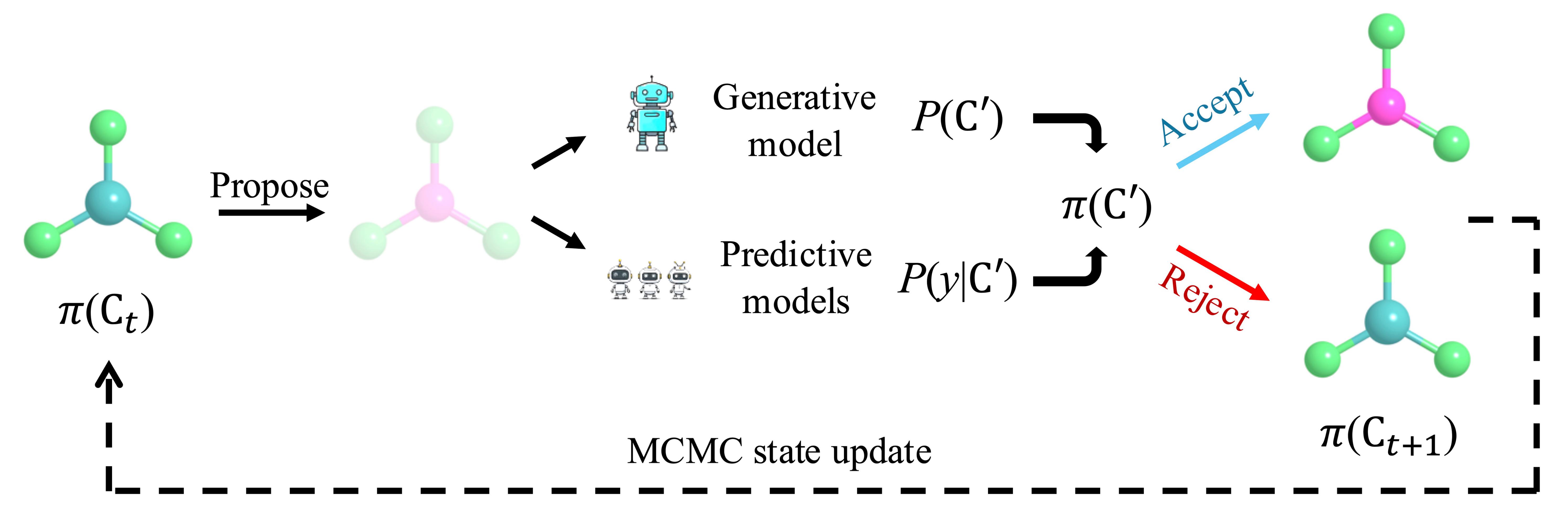}%
\caption{\label{fig:update} The fundamental steps of crystal structure conditional generation framework. This is one step of the Markov Chain Monte Carlo (MCMC) process, where $\pi(\cdot)$ represents the target distribution we aim to sample from. $\mathrm{C}_t$ denotes the state (crystal structure) at the t-th Markov step, and $\mathrm{C}'$ is the proposed updated state. $P\left(\mathrm{\cdot}\right)$ is the crystal probability given by the generative model, and $P\left(\mathrm{y|\cdot}\right)$ is the conditional probability provided by the prediction model, where y represents the target property we specify. The target distribution $\pi(\cdot)$ is determined by $P\left(\mathrm{\cdot}\right)$ and $P\left(\mathrm{y|\cdot}\right)$.}
\end{figure*}

\subsection{Conditional generation framework}

\subsubsection{Basic composition}

Most widely used general generative models in the field of crystal structure generation, such as autoregressive models, diffusion models, and flow-based models, are probabilistic generative models. Fundamentally, these models generate new crystal structures by learning the distribution of crystal structures present in the training dataset. Generating structures with these models can be understood as sampling from the distribution $P\left(\mathrm{C}\right)$, where $\mathrm{C}$ represents crystal structure and $P\left(\mathrm{C}\right)$ is the learned distribution approximating the true distribution $P^*\left(\mathrm{C}\right)$ observed in the training data. However, when aiming to generate crystals within a specific domain, the objective shifts to sampling from the conditional distribution $P^*\left(\mathrm{C}|y\right)$, where $y$ denotes the target properties characterizing the materials within that domain. And it is well knew that $P^*\left(\mathrm{C}|y\right)=P^*\left(C\right)P^*\left(y|\mathrm{C}\right)/P^*\left(y\right)$.
 Here, $P^*\left(y\right)$ acts as a normalization constant that can be ignored. Therefore, sampling from the distribution $P^*\left(\mathrm{C}|y\right)$ can be reformulated as sampling from the distribution $\pi^*\left(\mathrm{C}\right)=P^*\left(\mathrm{C}\right)P^*\left(y|\mathrm{C}\right)$.

Building upon the above analysis, we developed PODGen, a highly transferable and robust conditional generation framework. This framework consists of three key components: (1) a general generative model that provides $P\left(\mathrm{C}\right)$ to approximate $P^*\left(\mathrm{C}\right)$, (2) multiple predictive models that provide $P\left(y|\mathrm{C}\right)$ to approximate $P^*\left(y|\mathrm{C}\right)$, and (3) an efficient sampling method—Markov Chain Monte Carlo (MCMC) sampling. The fundamental steps of this framework are illustrated in FIG.~\ref{fig:update}.

In this framework, the generative model only needs to provide probabilistic estimates of crystal structures, without restrictions on the specific model type. Additionally, most widely used predictive models can be seamlessly integrated into this framework. For classification-based predictive models, which commonly employ cross-entropy loss, inherently yield probability estimates for each structural class; whereas for regression-based predictive models which commonly use the Mean Squared Error (MSE) as their loss function, which corresponds to a probabilistic model assuming that the observed data follow a Gaussian distribution centered around the predicted value with a fixed variance. Therefore, most predictive models can provide the probability $P\left(y|\mathrm{C}\right)$.

\subsubsection{Crystal generation}

MCMC sampling is an efficient method for sampling from complex high-dimensional distributions. Similar method has been used in language model for generating sentences that satisfy certain conditions\cite{miao2019cgmh,zhang2020language}, or optimizing the result\cite{song2025llmfeynmanleveraginglargelanguage}. MCMC generates a sequence of correlated samples by iteratively transitioning from one state to another based on the transition matrix of a Markov chain. In the context of crystal structure generation, each state corresponds to a specific crystal structure\cite{crystalformer}. The Metropolis-Hastings (MH) algorithm enables efficient computation of transition probabilities in a Markov chain\cite{HM_M,HM_H}. This algorithm proposes potential new states based on a designed update strategy and then accepts the transition with a probability given by Eq.~\ref{eq:MCMC_MH1} Eq.~\ref{eq:MCMC_MH2}. The detailed balance condition established in this way ensures that the samples obtained through MCMC conform to the target distribution.

\begin{eqnarray}
A\left(\mathrm{C}'|\mathrm{C}_{t-1}\right)=\mathrm{min}\left\{1, A^{*}\left(\mathrm{C}'|\mathrm{C}_{t-1}\right) \right\}
\label{eq:MCMC_MH1},
\\
A^{*}\left(\mathrm{C}'|\mathrm{C}_{t-1}\right) = \frac{\pi \left(\mathrm{C}'\right) P\left(\mathrm{C}_{t-1}|\mathrm{C}'\right)}{\pi \left(\mathrm{C}_{t-1}\right) P\left(\mathrm{C}'|\mathrm{C}_{t-1}\right)}
\label{eq:MCMC_MH2}.
\end{eqnarray}
Here, $\pi\left(\cdot \right)$ represents the target probability distribution, $P\left(\mathrm{C}'|\mathrm{C}_{t-1}\right)$ denotes the probability of proposing a transition from crystal structure $\mathrm{C}_{t-1}$ to crystal structure $\mathrm{C}'$, and $A\left(\mathrm{C}'|\mathrm{C}_{t-1}\right)$ is the acceptance probability for this proposed transition. If the proposal is accepted, then $\mathrm{C}_{t}$ = $\mathrm{C}'$; otherwise, $\mathrm{C}_{t}$ = $\mathrm{C}_{t-1}$. In this study, we applied this conditional generation framework to the generation of topological insulators, where $\pi\left(\mathrm{C} \right)$ is specifically defined as:

\begin{eqnarray}
\pi\left(\mathrm{C}\right) = P\left(\mathrm{C}\right)^{k}P\left(\mathrm{TI}|\mathrm{C}\right)P\left(\mathrm{NMet}|\mathrm{C}\right)P\left(\mathrm{NMag}|\mathrm{C}\right)\label{eq:target1},
\\
k=
\left\{
\begin{aligned}
& e^{0.5} , && \text{if } \exists e \in E , e \in C \\
& 1 , && \text{else}
\end{aligned}
\right. \label{eq:target2}.
\end{eqnarray}

In this paper we train a CrystalFormer\cite{crystalformer} as the general generation model to provide the $P\left(\mathrm{C}\right)$, and three classification model\cite{dimenet} to provide $P\left(y|\mathrm{C}\right)$ where TI stands for topological insulator, NMet stands for non-metal, and NMag stands for non-magnetic. Since existing rapid identification tools\cite{he2019symtopo, tqcmethod} for crystal topological properties are all symmetry-based, we selected CrystalFormer, which inherently encodes space group and Wyckoff position information within it. In contrast, the predictive models were not meticulously curated or extensively trained, further demonstrating the robustness of our framework. For more information about these model, please refer to Supplementary Section S1. And thought the analysis of the database\cite{zhang2019catalogue,tqc}, it is not difficult to find that topological insulators are more likely to be found in crystals containing these elements $E=\left\{\mathrm{Al, P, S, Ga, Ge, As, Se, In, Sn, Sb, Te, Pb, Bi}\right\}$. Therefore, we introduce $k$ to modify the crystal probability $P\left(\mathrm{C}\right)$, aiming to generate crystals containing these elements with a higher probability.

We employ three types of proposals, each corresponding to modifications in atomic species, atomic coordinates, and lattice constants. At each step of the Markov chain, one of these proposals is randomly selected with probabilities of 0.2, 0.4, and 0.4, respectively. When modifying atomic species, we first select a Wyckoff position from the current configuration with equal probability and then replace the atomic species at that position with a randomly chosen element. For atomic coordinate modifications, we apply Gaussian noise to the fractional coordinates of all atoms with degrees of freedom, while respecting Wyckoff position constraints, which may prevent certain atomic coordinates from being altered. Similarly, when modifying lattice constants, Gaussian noise is added to all adjustable lattice parameters, subject to space group constraints, which may restrict changes to certain lattice constants.

\begin{figure}[h]
\includegraphics[width=0.45\textwidth]{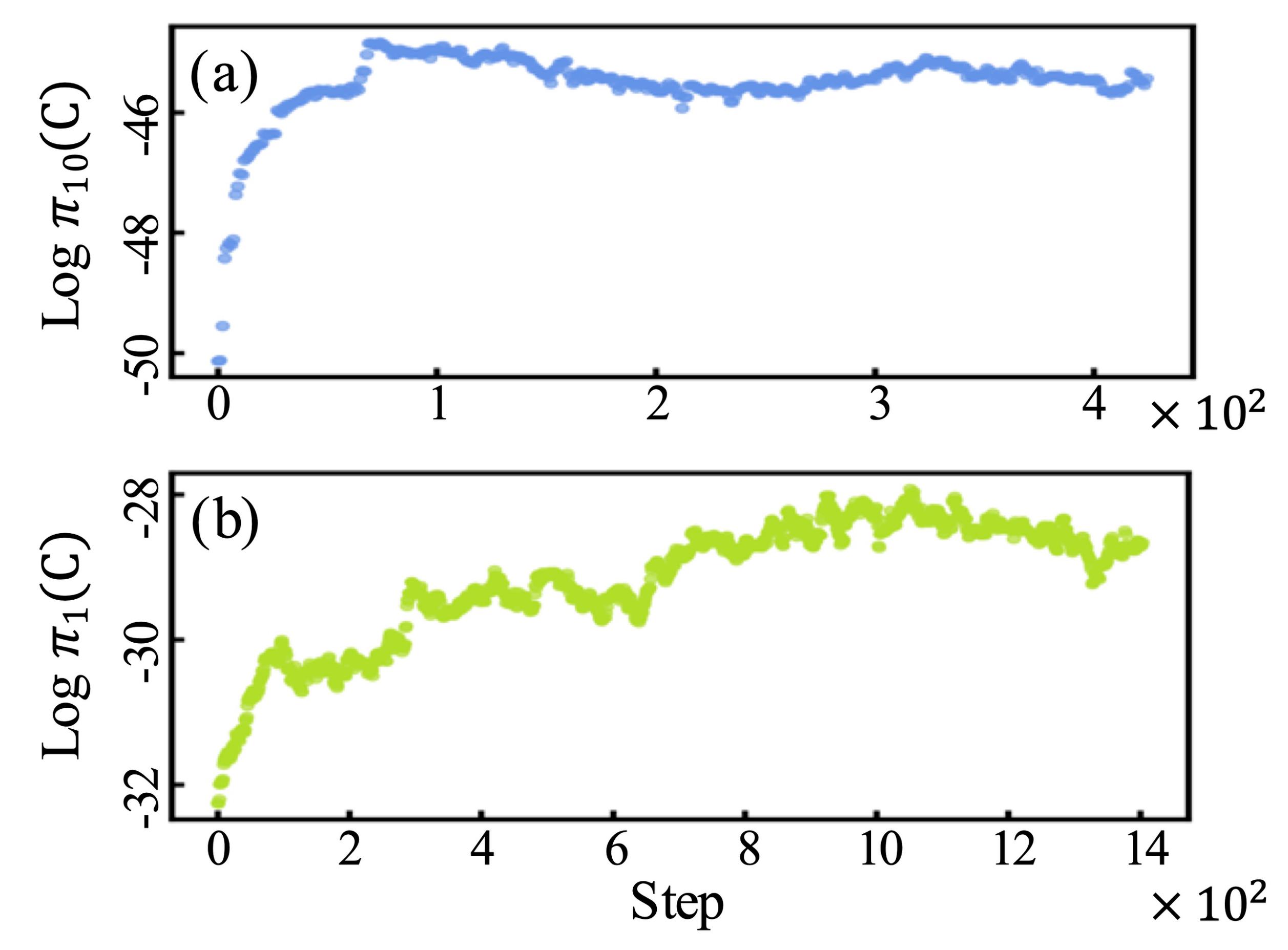}%
\caption{\label{fig:mcmc} One of the annealing MCMC convergence curves is presented, where the horizontal axis represents the Markov step count, and the vertical axis denotes the logarithmic probability of the target distribution at the corresponding temperature. (a) The highest temperature T = 10. (b) The lowest temperature T = 1.}
\end{figure}

To prevent the generated structures from being confined to known regions of configuration space, we incorporate a simulated annealing approach with ten temperature levels ranging from T = 10 to T = 1. The process begins by randomly selecting a crystal structure from the Alexand20\cite{schmidt2022dataset,schmidt2022large} (refer to Supplementary Section S1) as the initial state. Starting from T = 10, we allow the system to equilibrate at each temperature before gradually cooling to the next level, continuing until convergence is reached at T = 1, at which point sampling is performed. Convergence at each temperature is determined based on the rolling window mean and standard deviation of $\mathrm{Log}\ \pi_{T}\left(\mathrm{C}\right)$, with a window size of 200 steps and a tolerance of 1e-3. Fig.~\ref{fig:mcmc} presents the convergence curves at T = 10 and T = 1. During the sampling phase, we record a sample every 100 Markov steps.

\subsection{Crystal generation workflow}

\begin{figure*}[t]
\includegraphics[width=1.0\textwidth]{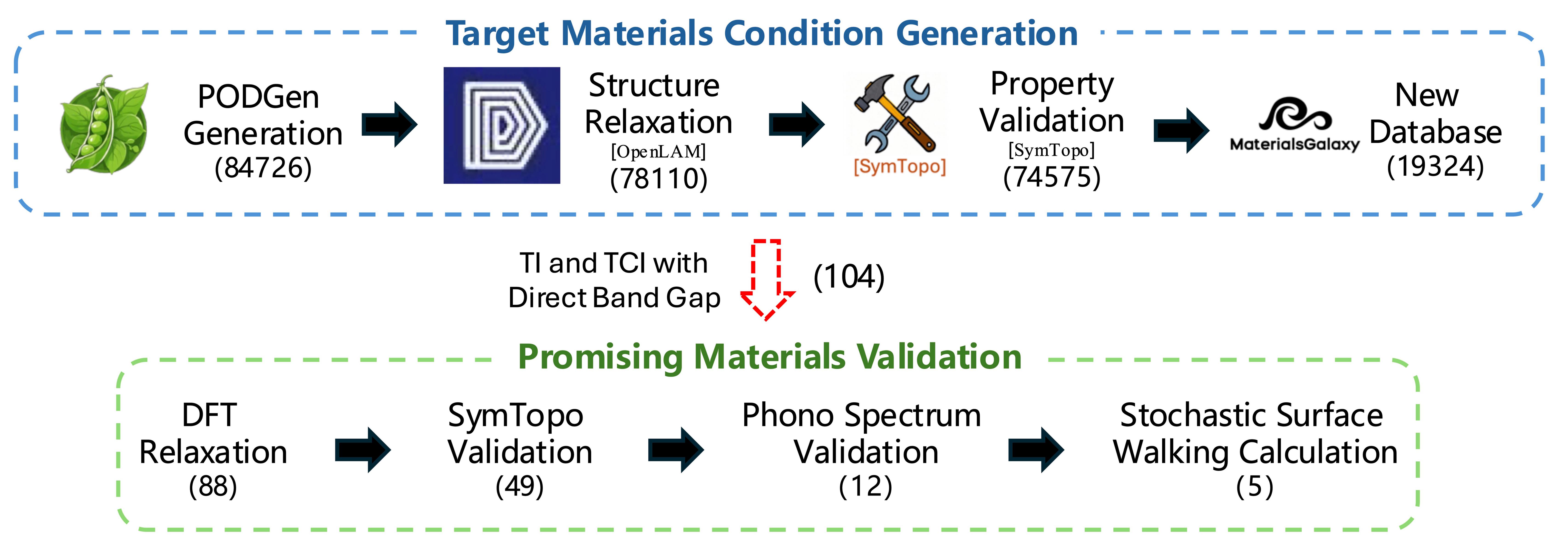}%
\caption{\label{fig:workflow} The core workflow of this study involves utilizing a conditional generation framework, which incorporates a general generative model, to generate crystal structures. Subsequently, various tools are employed for structural optimization, screening, and evaluation. Finally, materials with promising application potential undergo further first-principles calculations for validation.}
\end{figure*}

We have designed a workflow for high-throughput generation of crystal structures in specific domains, as illustrated in FIG.~\ref{fig:workflow}. This workflow encompasses conditional crystal structure generation, machine learning force field relaxation, crystal property evaluation, and structure deduplication. We applied this workflow to the generation of topological insulators and further validated the most promising candidate materials through first-principles calculations.

The workflow integrates our conditional generation framework PODGen, a general machine learning force field (MLFF) OpenLAM\cite{peng2025openlam}, a symmetry-based topological classification tool Symtopo\cite{he2019symtopo}, and first-principles calculation tools such as VASP\cite{kresse1996efficiency}, along with software packages including pymatgen\cite{ong2013python}, ASE\cite{larsen2017atomic}, VASPkit\cite{wang2021vaspkit}, and Phonopy\cite{phonopy-phono3py-JPCM,phonopy-phono3py-JPSJ}. This workflow is transferable to the exploration of other condition-dependent crystal materials by simply replacing the corresponding prediction models and crystal property evaluation tools.

In this high-throughput generation workflow we first use MLFF model to relax generated structure. The MLFF relaxation closely approximates first-principles results while being approximately three orders of magnitude more efficient. And we employed the OpenLAM model released in October 2024\cite{peng2025openlam}. Then we use SymTopo\cite{he2019symtopo} to help us quickly verify the topological properties of new crystals. SymTopo is an automated tool for calculating the topological properties of nonmagnetic crystalline materials. At last, we use $StructureMatcher$ module from pymatgen\cite{ong2013python} to determine whether two structures are similar. The scope of duplicate checking is the combined dataset\cite{heyu} of Materiae\cite{zhang2019catalogue} and TQC\cite{tqc}, as well as the newly generated crystals. For more details, please refer to Supplementary Section S2.

\subsection{Promising Crystals Validation}

For structures with a direct band gap that are classified as topological insulators or topological crystalline insulators after MLFF relaxation, we further verify their properties and stability using first-principles calculations. This process includes DFT relaxation, reconfirmation of topological classification using SymTopo, and phonon spectrum analysis. To ensure computational efficiency and reproducibility, we primarily use pymatgen and VASPkit to generate input files for VASP, with all calculations performed using VASP 5.4.4.

During the relaxation process, we divide it into two steps. In the first step, we use VASP input files generated by pymatgen for relaxation. However, since the default convergence criteria in pymatgen are not stringent enough (EDIFF is typically set on the order of 1e-3), we refine the relaxation in a second step. Once the first relaxation succeeds, we modify EDIFF to 1e-5 and EDIFFG to -1e-3 in the second step, ensuring that the atomic forces are reduced to below 1e-3 eV/\AA.

If both relaxation steps converge, we use SymTopo to reassess the topological properties and band gap of the relaxed structure. For structures that remain topological insulators with a direct band gap, we further compute their electronic band structure and phonon spectrum. For band structure calculations, we directly use the CHGCAR and fermi energy obtained from SymTopo’s SCF calculation, along with the high-symmetry path generated by VASPkit, to plot the band structure. For phonon spectrum calculations, we employ a 2×2×2 supercell and the DFPT method. Except for setting ENCUT to 1.3 times ENMAX, all other parameters are generated by VASPkit. Most DFT calculations in this paper use the PBE exchange-correlation functional\cite{PBE}, and some of the results obtained using the HSE06 functional\cite{hse06} can be found in the Supplementary Information.

To further evaluate the synthesizability of these structures, we employed the Stochastic Surface Walking (SSW) method\cite{ssw,sswcrystal} to explore their potential energy surfaces. This not only allowed us to determine the locations of the generated crystal structures on the PES, but also enabled the discovery of more stable configurations.

\section{Result}

\subsection{Topological Material Condition Generation}

\begin{figure*}[t]
\includegraphics[width=1.0\textwidth]{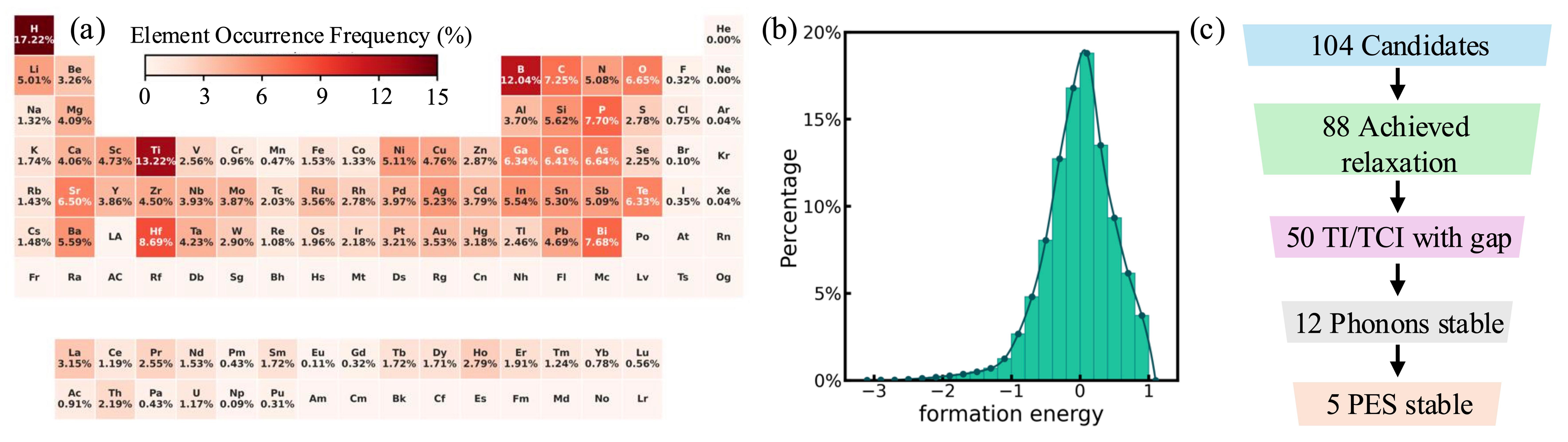}%
\caption{\label{fig:statis} (a) The elemental occurrence frequency distribution among the generated TI and TCI materials, and (b) The formation energy distribution predicted by OpenLAM. (c) Further calculations and screening performed on 104 TI and TCI materials with direct band gaps.}
\end{figure*}

\begin{table}[h]
\caption{\label{tab:table2}%
The proportions of topological insulators and topological crystalline insulators, as well as their ratio, in both general generation and conditional generation using topological insulators as constraint.
}
\begin{ruledtabular}
\begin{tabular}{cccc}
\textrm{Method}&
\textrm{TI}&
\textrm{TCI}&
\textrm{TI:TCI}\\
\colrule
General generation & 2.85\% & 2.45\% & 1.16:1 \\
Conditional generation & 15.25\% & 9.93\% & 1.62:1 \\
\end{tabular}
\end{ruledtabular}
\end{table}

Using the method mentioned before, we have generated 84726 crystal, 78110 of them can be successfully relaxed by OpenLAM with maximum atomic force falls below 0.02 eV/\AA \ and predicted formation energy smaller then 1.0 eV/atom. Then 78575 of them can get the topological classification given by Symtopo. After removing the duplicate structures, there are 11914 unique crystals classified as TI and 7336 unique crystals classified as TCI, corresponding to proportions of 15.25\% and 9.93\%. Here, 68 TIs are considered to have direct band gaps, among which 63 are also regarded as having indirect band gaps. Among the 36 TCIs considered to have direct band gaps, 34 are also regarded as having indirect band gaps.

We also explored the direct generation of crystal structures using CrystalFormer for topological material screening. Among the 2000 generated materials, only 57 were identified as TI and 49 as TCI, corresponding to probabilities of 2.85\% and 2.45\%, respectively. This generation efficiency is significantly lower than that achieved through conditional generation. More importantly, no gapped TIs were found among them. Furthermore, we observed that in the absence of conditional generation, the ratio of TI to TCI was 1.16:1. However, when generating materials conditioned on TI (excluding TCI), this ratio increased to 1.62:1. As shown in Supplementary Table S1, the topological classification model we employed is a relatively basic one, and TI and TCI are known to be categories that are prone to misclassification by predictive models. Nevertheless, our conditional generation framework significantly improves both the success rate and the proportion of materials with the desired topological properties, demonstrating its robustness. We believe that employing a state-of-the-art (SOTA) predictive model with refined training will further enhance generation efficiency.

We conducted a statistical analysis of the 19,250 generated TI and TCI materials. FIG.~\ref{fig:statis}(a) shows the occurrence frequency of each element in these materials. Compared to the CrystalFormer training set (Supplementary Fig.~S2), the elemental distribution of the generated crystals has been substantially altered, resembling more closely the distribution of topological insulators in the existing database\cite{heyu} (Supplementary Fig.~S4). This indicates that our conditional generation framework effectively adjusts the baseline distribution of CrystalFormer toward the target distribution characteristic of topological materials. Further analysis reveals that, although elements such as B and Ge maintain high occurrence frequencies similar to those in the existing database, the most prevalent element shifted from O in the original database to H in the newly generated materials. This shift demonstrates that our framework not only aligns with the existing distribution but also explores new compositional spaces beyond the limitations of the original datasets, leading to the discovery of novel crystal structures.

FIG.~\ref{fig:statis}(b) presents the formation energy distribution of the 19,250 generated TI and TCI materials, with formation energies predicted concurrently during relaxation using OpenLAM. Although the distribution does not fully align with the ideal scenario where all formation energies are negative, it closely resembles the formation energy distribution of the CrystalFormer training set (Supplementary Fig.~S3). This observation suggests that for properties not explicitly constrained by our conditional generation framework—such as formation energy, which has little direct correlation with topological properties—the generated crystal structures largely adhere to the inherent distribution of the base model.

\subsection{Promising Materials Validation}

\begin{figure*}[t]
\includegraphics[width=1.0\textwidth]{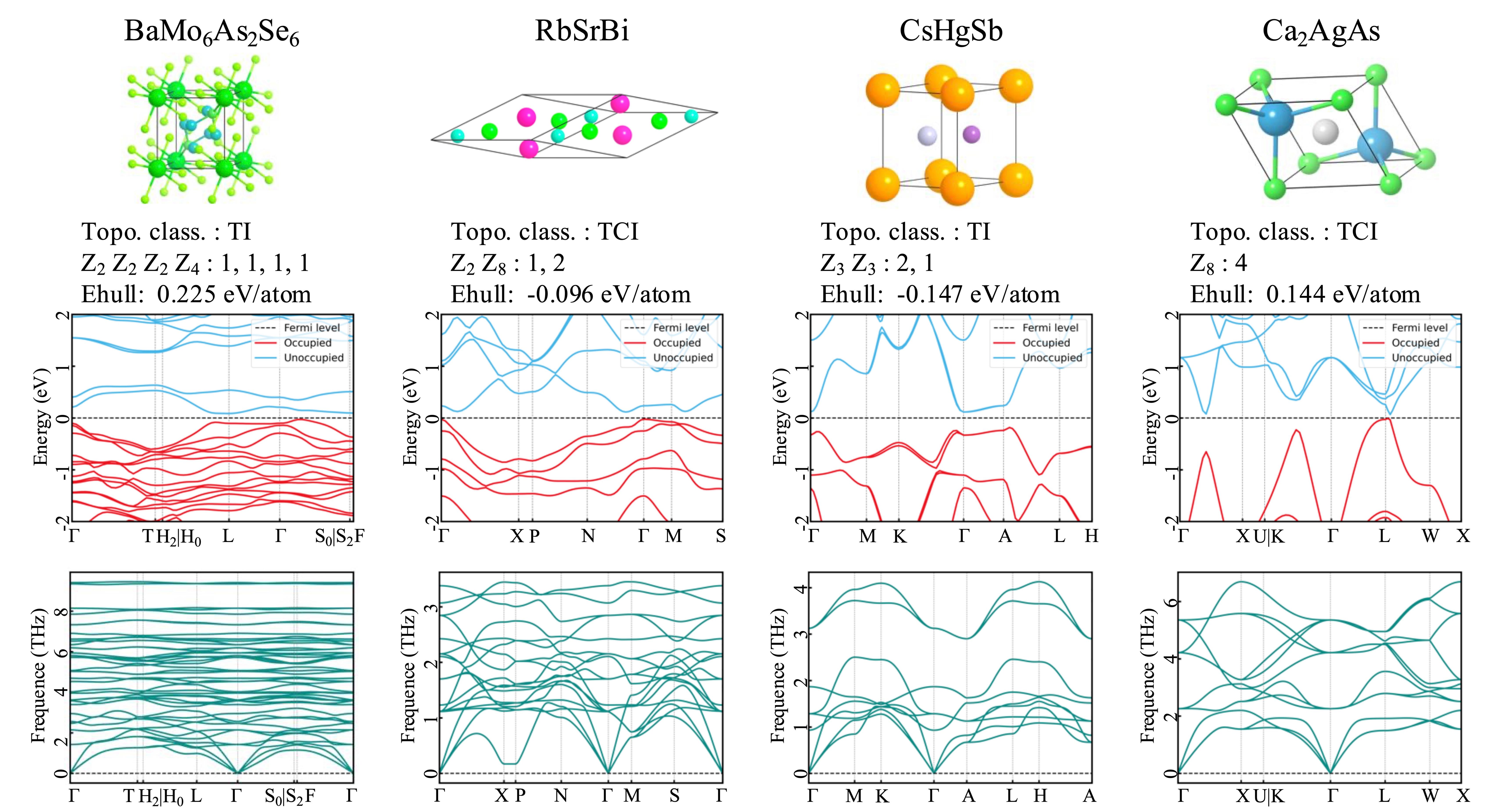}%
\caption{\label{fig:Crystal_bs_phonon} Four crystal structures that remain topological (Crystalline) insulators with a direct band gap after further relaxation via DFT, and exhibit phonon spectra without imaginary frequencies as calculated by DFPT. The topological classification and symmetry-based indicators for these structures are provided by SymTopo\cite{he2019symtopo}. Ehull represents the energy above hull, which are obtained from DFT calculations.}
\end{figure*}

Although materials without a band gap can still be identified as TI through symmetry-based topological classification methods\cite{he2019symtopo,tqcmethod}, gapped TI are generally considered more promising for practical applications. Therefore, from the generated TI and TCI materials, we selected 104 candidates with a direct band gap for further validation through first-principles calculations. The verification process is outlined in FIG.\ref{fig:statis}(c). Among these materials, 88 were successfully relaxed, and 50 retained their direct band gap as TI or TCI after relaxation. Notably, 12 of these materials exhibited phonon spectra without imaginary frequencies. The crystal structures, SymTopo classification results, topological indices, band structures, and phonon spectra of these 12 materials are presented in FIG.~\ref{fig:Crystal_bs_phonon} and Supplementary Fig.~S6.

\begin{figure}[h]
\includegraphics[width=0.5\textwidth]{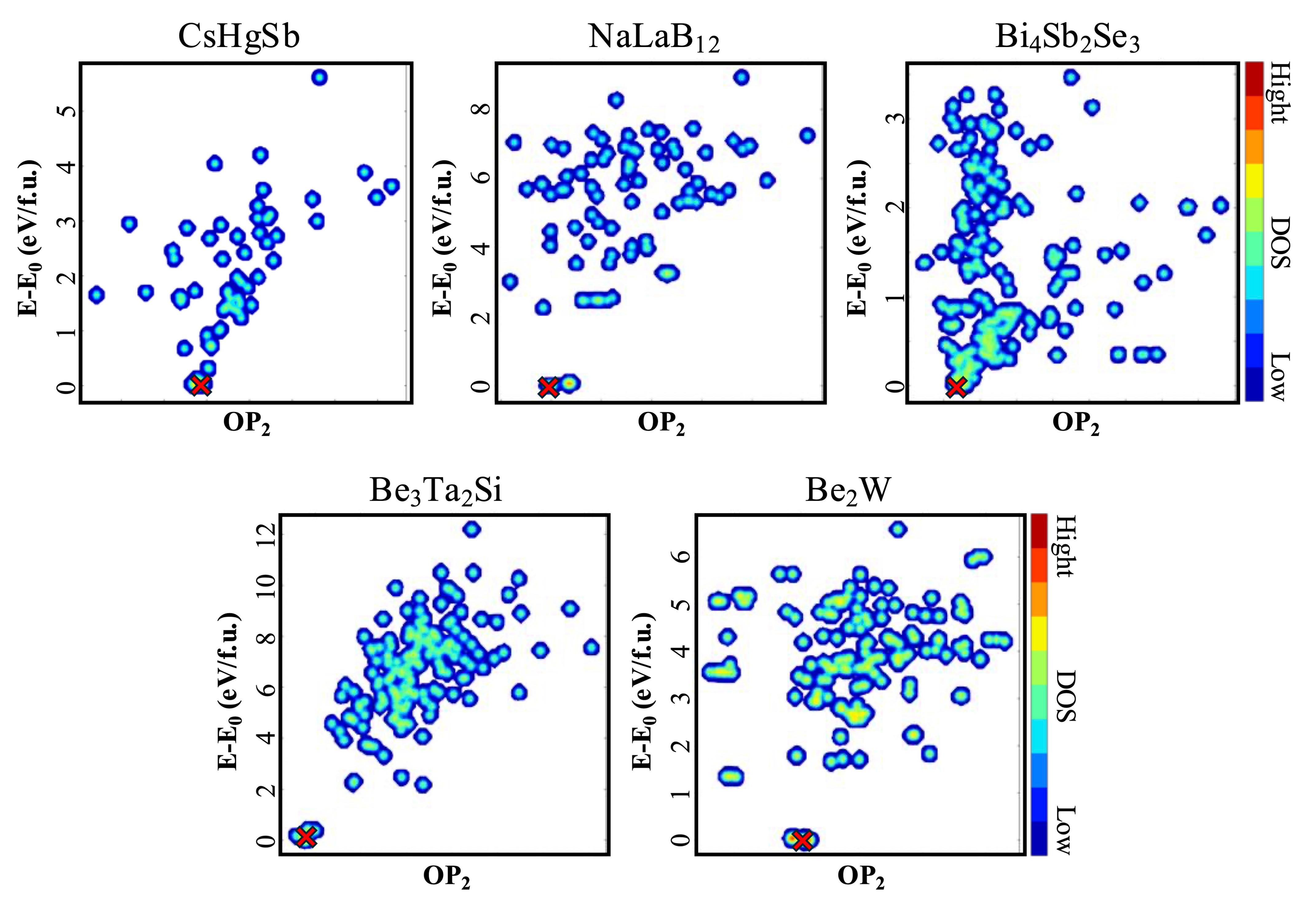}  
\caption{\label{fig:pesmain} PES contour plots of five newly generated materials obtained from SSW, where the vertical axis represents the relative energy and the horizontal axis denotes the structural descriptor. ‘DOS’ indicates the number of times each structure was found. The red crosses mark the positions of these five materials on their respective potential energy surfaces.}
\end{figure}

We applied the SSW method to explore the PES of this 12 materials, in order to further evaluate their experimental synthesizability. As shown in FIG.~\ref{fig:pesmain}, these five materials are located at the bottom of their PES and exhibit negative energy above hull (as shown in FIG.~\ref{fig:Crystal_bs_phonon} and Supplementary Fig.~S6), indicating a higher likelihood of experimental realization. The PES landscapes of the remaining materials are shown in Supplementary Fig.~S7.

\subsection{WannierTools confirmation}

\begin{figure*}[t]
\includegraphics[width=1.0\textwidth]{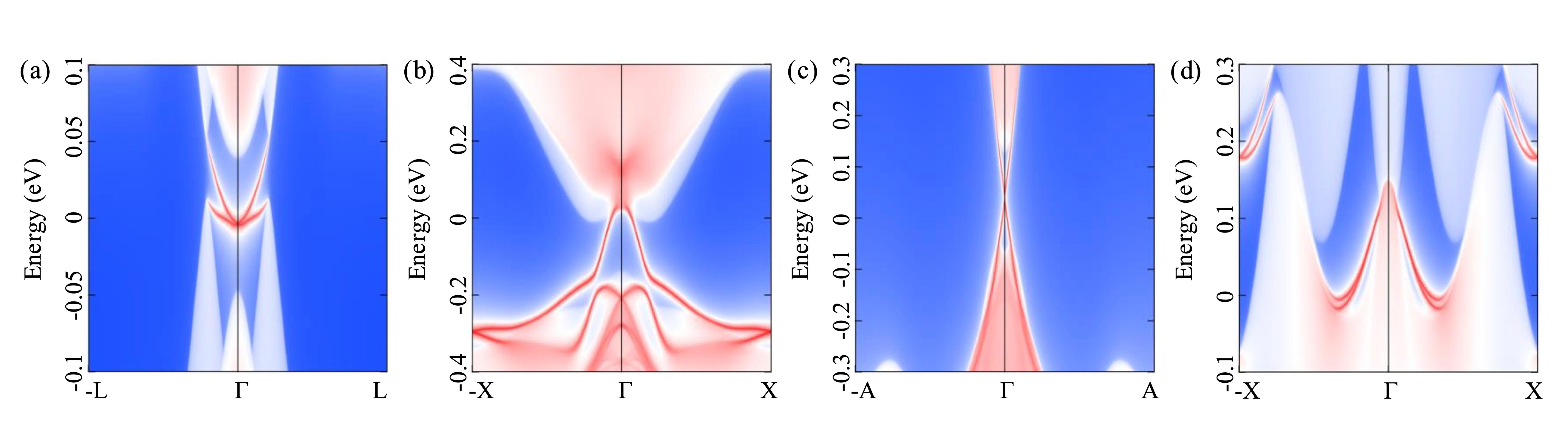}  
\caption{\label{fig:surface} Boundary-state spectra for three materials, each plotted along the indicated high-symmetry lines. (a) $\mathrm{BaMo_6As_2Se_6
}$ with an open boundary on the (010) surface, (b) $\mathrm{RbSrBi}$ with an open boundary on the (100) surface, (c) $\mathrm{CsHgSb}$ with open boundary on the (100) surface, and (d) $\mathrm{Ca_2AgAs}$ with an open boundary on the (010) surface. The red curves highlight the in-gap boundary modes.}
\end{figure*}

To further validate our results, we selected several materials and constructed Wannier tight-binding models\cite{tight1,tight2,tight3} using Wannier90\cite{wannier90}. We then used the WannierTools\cite{wu2018wanniertools} package to calculate surface states and the Wilson loops\cite{z21,z22,yu2011equivalent} based on these Wannier models. In our analysis, open boundary conditions were imposed along different crystallographic directions: FIG.~\ref{fig:surface}(a) presents the boundary-state spectrum of $\mathrm{BaMo_6As_2Se_6}$, with an open boundary on the (010) surfac. whereas FIG.~\ref{fig:surface}(b) and FIG.~\ref{fig:surface}(c) show the spectra of $\mathrm{RbSrBi}$ and $\mathrm{CsHgSb}$, respectively, with an open boundary on the (100) surface, and FIG.~\ref{fig:surface}(d) displays $\mathrm{Ca_2AgAs}$ with an open boundary on the (010) surface. As illustrated in the figures, pronounced in-gap boundary states are clearly observed, indicating that these materials are topological (Crystalline) insulators. Detailed Wilson loop calculations can be found in Supplementary Fig.~S5.

\section{Conclusion}

Crystal structure generation models are powerful tools for discovering novel crystalline materials. However, when searching for structures with specific properties, conditional generation methods can significantly enhance efficiency. In this study, we developed a highly transferable and robust conditional generation framework PODGen by integrating the general crystal generation model, with multiple property prediction models. A generative model capable of providing crystal structure probabilities and most existing predictive models can be seamlessly incorporated into this framework, which imposes minimal requirements on their predictive capabilities. Moreover, once the base model is trained, conditional generation can be performed simply by training an appropriate predictive model for any specific domain. This significantly reduces the dependence on large domain-specific datasets and lowers training costs.

For properties explicitly conditioned by a predictive model (or those with strong correlations), our approach effectively guides the base model—originally trained to follow the distribution of the training dataset—toward generating structures that conform to a desired target distribution. Conversely, for properties without an associated predictive model (or those with weak correlations), the generated structures continue to follow the original training distribution.

We applied this framework to the conditional generation of topological insulator materials, achieving a success rate 5.35 times higher than that of conventional generation models. More importantly, the stricter the property constraints on the generated crystals, the greater the advantage of our framework over general generative models. For example, in generating gapped  TIs, our framework achieves success where general methods almost entirely fail—representing an effectively $\infty$ improvement. Using this method, we generated over 80,000 structures, nearly 20,000 of which were identified as TI or TCI. Further first-principles calculations were performed on the subset with direct band gaps, leading to the identification of 12 materials with promising application potential. Five of these structures are located near the global minima of the PES, suggesting a higher likelihood of experimental synthesis. Furthermore, we used WannierTools to further verify our results.

Certainly, there remains room for improvement in our framework. We adopted CrystalFormer as the base model; however, during the MCMC state updates, we only modified atomic species, atomic positions, and lattice constants, while leaving Wyckoff positions and space groups unchanged. This limitation arises because, in the string-based crystal structure representation used by CrystalFormer, Wyckoff positions are interdependent, requiring a more sophisticated update strategy. Additionally, modifications to the space group would fundamentally alter the entire structural representation of CrystalFormer.

\section{Code and Data}
Our code is available at \url{https://github.com/cyye001/PODGen}. And the dataset of generated crystals will be shown on a website of the Condensed Matter Physics Data Center of Chinese Academy of Sciences \url{https://cmpdc.iphy.ac.cn/materialsgalaxy/#/services/materials} and can be downloaded in Electronic Laboratory for Material Science \url{https://in.iphy.ac.cn/eln/link.html#/113/G9f5}.


~\\
\sect{Acknowledgements}
\noindent We thank Shigang Ou, Ruihan Zhang, Jingyu Yao, Yue Xie, Yi Yan, Yuanchen Shen for useful discussions. This work was supported by the Science Center of the National Natural Science Foundation of China (Grant No. 12188101), the National Key Research and Development Program of China (Grant No. 2023YFA1607400, 2022YFA1403800), the National Natural Science Foundation of China (Grant No.12274436, 11921004), and H.W. acknowledge support from the New Cornerstone Science Foundation through the XPLORER PRIZE.
~\\

\sect{Author contributions}
\noindent C.Y., H.W. and Q.W. conceived the idea and performed the analysis. C.Y. developed and implemented the PODGen framework, designed and executed the crystal generation workflow, and conducted the DFT, SymTopo, and DFPT calculations. Y.W. performed the WannierTools calculations. X.X. and Z.L. carried out the SSW calculations. T.Z. presented the results on the website. Y.H. provided the topological materials dataset and performed data cleaning. All authors contributed to the interpretation of the results and the writing of the manuscript.

~\\

\sect{Competing interests}
\noindent The authors declare no competing interests.\par 
~\\
\sect{Additional information}\par
\subsect{Supplementary information}
~\\
\textbf{Correspondence} and requests for materials should be addressed to Quansheng Wu.
\newpage
\clearpage

       \beginsupplement{}
        \setcounter{section}{0}
        \renewcommand{\thesubsection}{S\arabic{subsection}}
        \renewcommand{\thesubsubsection}{\Alph{subsubsection}}
        \titleformat{\section}[hang]{\bf\centering\Large}{\section}{1.0em}{}[]
        \titleformat{\subsection}[hang]{\bf\large}{\thesubsection}{0.5em}{}[]
        \titleformat{\subsubsection}[hang]{\bf\normalsize}{\thesubsubsection}{0.5em}{}[]

\end{document}


\soulregister{\cite}7
\soulregister{\ref}7
\tolerance 10000

\draft

       \beginsupplement{}
        \setcounter{section}{0}
        \renewcommand{\thesubsection}{S\arabic{subsection}}
        \renewcommand{\thesubsubsection}{\Alph{subsubsection}}
        \titleformat{\section}[hang]{\bf\centering\Large}{\section}{1.0em}{}[]
        \titleformat{\subsection}[hang]{\bf\large}{\thesubsection}{0.5em}{}[]
        \titleformat{\subsubsection}[hang]{\bf\normalsize}{\thesubsubsection}{0.5em}{}[]

\section*{Supplemental Materials}

\subsection{Models training}
\label{sec:models_training}

\subsubsection{CrystalFormer}

\begin{figure}[h]
\includegraphics[width=0.45\textwidth]{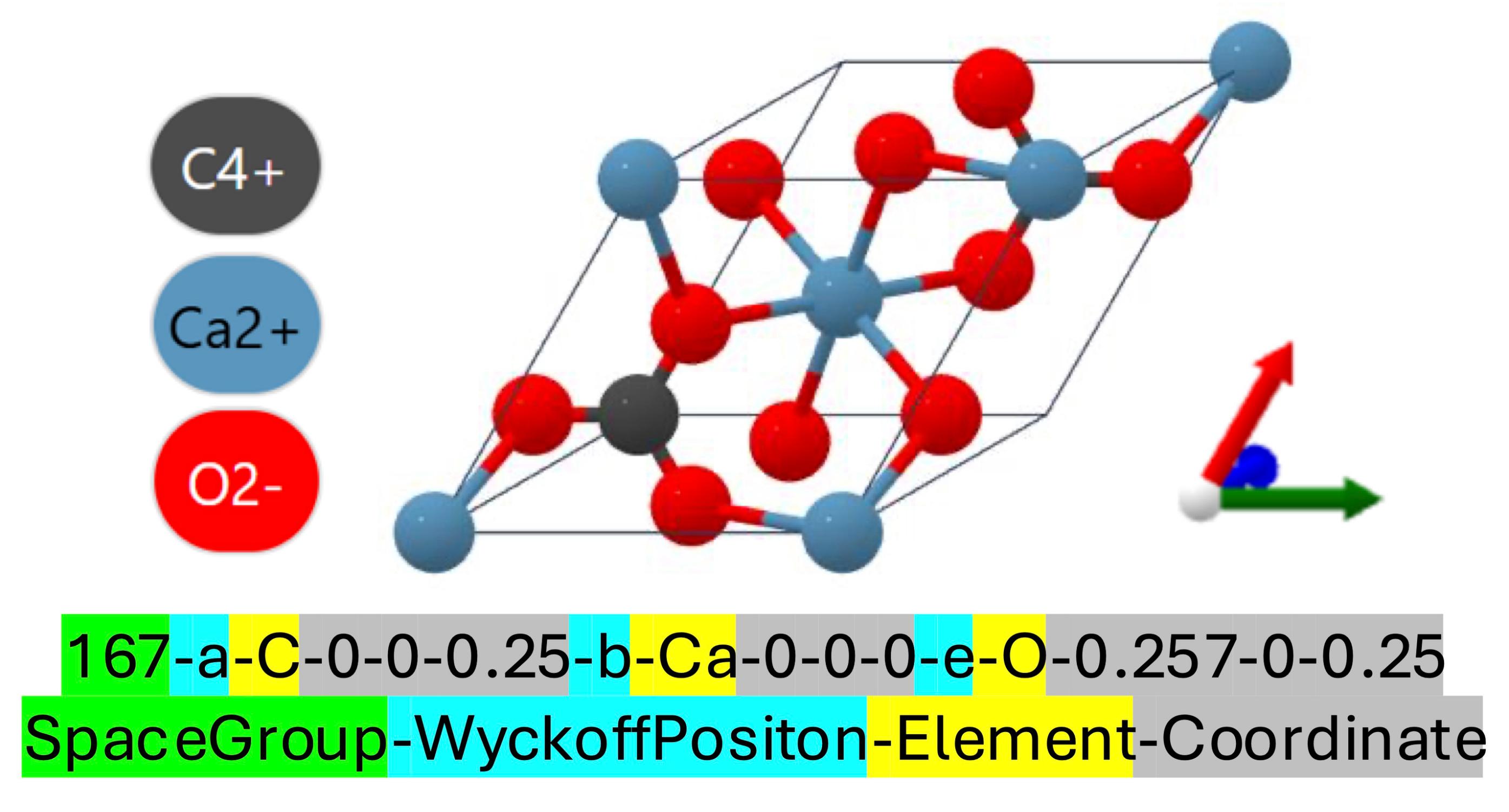}%
\caption{\label{fig:CrystalFormer} String-Like representation of crystal structures use by CrystalFormer\cite{crystalformer}.}
\end{figure}

CrystalFormer is a kind of autoregressive crystal structure generation model. Using the space group, wyckoff position and other information, CrystalFormer can convert crystal structures into a string-like format losslessly (as showed in FIG.~\ref{fig:CrystalFormer}), and generate crystal structures like a language model generates sentences. 

In this paper, we trained CrystalFormer using crystal structures from the Alexandria\cite{schmidt2022dataset,schmidt2022large} with unit cells containing no more than 20 atoms and formation energy smaller then 0.5 eV/atom, totaling approximately 1.8 million structures. In this paper, we refer to it as Alex, which was randomly split into training, test, and validation sets in an 8:1:1 ratio. This sufficiently large dataset enables the model to learn a broader and more diverse distribution of crystal structures which serves as the fundamental basis for this method.

\begin{figure}[h]
\includegraphics[width=0.45\textwidth]{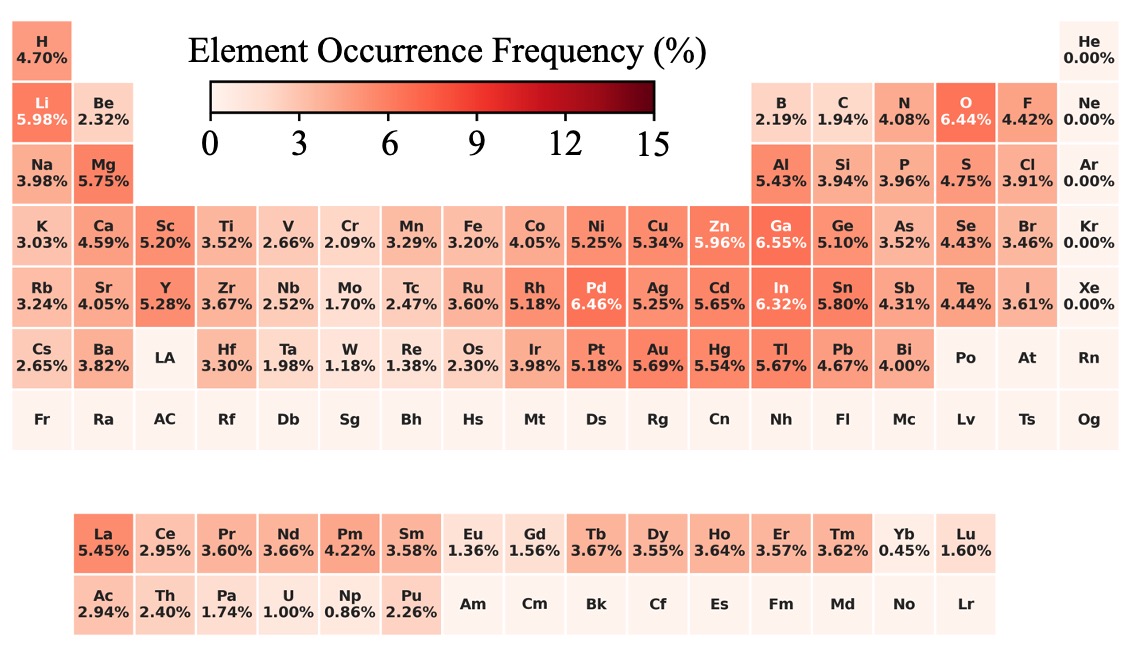}%
\caption{\label{fig:Alex_ele} The elemental occurrence frequency distribution in the Alexand dataset.}
\end{figure}

\begin{figure}[h]
\includegraphics[width=0.45\textwidth]{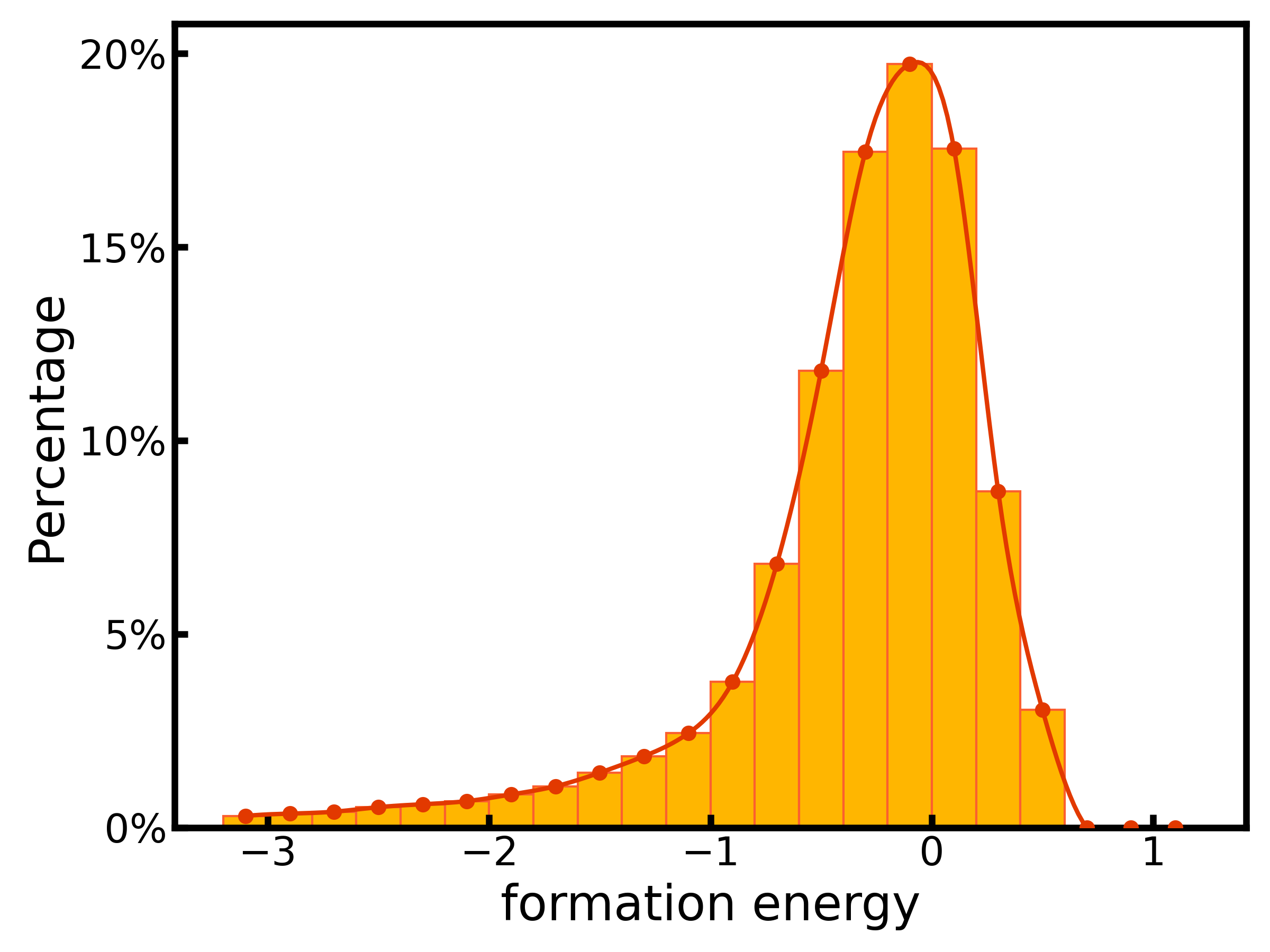}%
\caption{\label{fig:Alex_formation} The formation energy distribution within the Alexand20 dataset. }
\end{figure}

FIG.~\ref{fig:Alex_ele} and FIG.~\ref{fig:Alex_formation} illustrate the elemental distribution and formation energy distribution within the training dataset, Alexand20\cite{schmidt2022dataset,schmidt2022large}, of CrystalFormer\cite{crystalformer} in this paper. This dataset contains was specifically designed to ensure a broad and uniform distribution of elements and structural configurations, making it a strong foundation for training base model of crystal generation. However, this broad coverage comes at the expense of the formation energy distribution, as a significant portion of the materials exhibit positive formation energies, which is less ideal for practical applications.

\begin{figure}[h]
\includegraphics[width=0.45\textwidth]{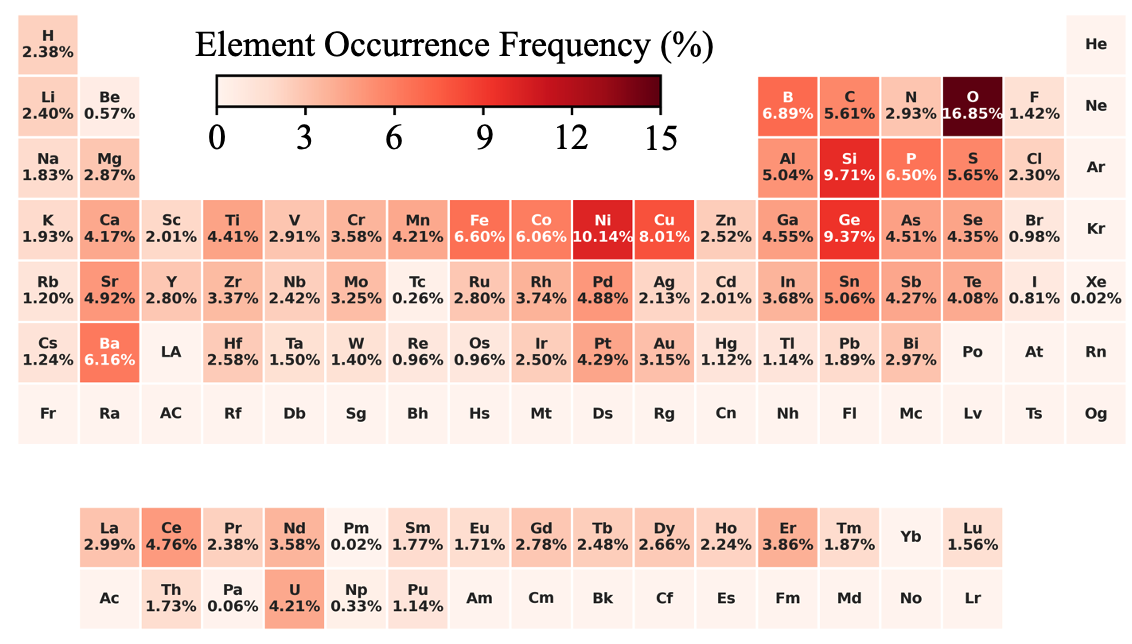}%
\caption{\label{fig:materiae_tqd_TITCI} The elemental occurrence frequency distribution among TI and TCI materials in the merged database combining the Materiae\cite{zhang2019catalogue} and TQC databases\cite{tqc}.}
\end{figure}

FIG.~\ref{fig:materiae_tqd_TITCI} shows the element occurrence frequency in topological (crystalline) insulators from the original database. It can be observed that this distribution differs significantly from that of the Alexand20 dataset.

\subsubsection{Property Prediction Models}

\begin{table}[h]
\caption{\label{tab:table1}%
The three classification models used in this paper.
}
\begin{ruledtabular}
\begin{tabular}{ccc}
\textrm{Property}&
\textrm{Number of categories}&
\textrm{Accuracy}\\
\colrule
Metal & 2 & 88.5\% \\
Magnetism & 2 & 93.1\% \\
Topological classification & 5 & 69.5\% \\
\end{tabular}
\end{ruledtabular}
\end{table}

For convenience and to demonstrate the robustness of our approach, we did not specifically select or design property prediction models. Instead, we uniformly adopted the DimeNet++\cite{dimenet} architecture to train three classification models: one for determining whether a material is metallic (i.e., whether it has a band gap), one for predicting magnetic properties, and one for topological classification. The training results are presented in Table~\ref{tab:table1}. The first two classification models were trained using data from the Materials Project\cite{MP}, while the topological classification model was trained using Materiae\cite{zhang2019catalogue} databases. It can be observed that the performance of these three models is not state-of-the-art (SOTA). However, as demonstrated in subsequent tests, they are already sufficient to enable our conditional generation framework to function effectively.

The trained classification models can, given a crystal structure, provide the probability of it being classified into each category, denoted as $P\left(y|\mathrm{C}\right)$. Here, $y$ represents the target properties we set. In this study, these correspond to non-metallic (NMet), non-magnetic (NMag), and topological insulator (TI). Although we only utilized classification models, regression-based predictive models can also be incorporated into this conditional generation framework. This is because most regression models use the Mean Squared Error (MSE) as their loss function, which corresponds to a probabilistic model assuming that the observed data follow a Gaussian distribution centered around the predicted value with a fixed variance. As a result, these models can also provide a probability $P\left(y|\mathrm{C}\right)$.

\subsection{Details of the high-throughput generation workflow}
\label{sec:details}

\subsubsection{MLFF Relaxation}

Unlike conventional predictive machine learning models, the MLFF model is trained on first-principles calculations of both equilibrium and non-equilibrium structures. This enables it to accurately predict the energy and atomic forces of various configurations, allowing for crystal structure relaxation that closely approximates first-principles results while being approximately three orders of magnitude more efficient.

For structural relaxation, we employed the model released by OpenLAM in October 2024\cite{peng2025openlam}. The relaxation process was carried out using the Broyden-Fletcher-Goldfarb-Shanno (BFGS) algorithm, with a maximum of 400 relaxation steps. Convergence was determined based on the criterion that the maximum atomic force falls below 0.02 eV/\AA. Using this model, we can also get the predicted energy of each structure, which is use to calculate the formation energy, and we filter out the crystals with formation energy greater then 1.0 eV/atom.

\subsubsection{Topological Property Determination}

We use SymTopo\cite{he2019symtopo} to help us quickly verify the topological properties of new crystals. SymTopo is an automated tool for calculating the topological properties of nonmagnetic crystalline materials. Based on topological quantum chemistry and symmetry indicator theory, it significantly simplifies the process of determining material topology. The tool employs symmetry representation methods and integrates with first-principles calculation software such as VASP and ABINIT to automatically analyze the electronic band structures and predict topological properties. SymTopo is primarily used for screening topological materials, identifying different categories such as topological insulators and topological semimetals, and assisting researchers in efficiently discovering new topological materials. Its algorithm includes electron count checking, space group identification, density functional theory (DFT) calculations, band degeneracy analysis, and symmetry indicator computation.

We performed the necessary DFT calculations for SymTopo using VASP 5.4.4\cite{vasp}, with input files generated by SymTopo. Two key parameters were modified. By default, SymTopo does not specify ENCUT, which causes VASP to adopt the maximum ENMAX from the POTCAR file. To enhance computational accuracy, we set ENCUT to 1.3 times the maximum ENMAX. Additionally, SymTopo sets ISMEAR = -5 by default, corresponding to the Tetrahedron method for determining partial occupancies. However, this setting can lead to errors with certain KPOINTS configurations generated by SymTopo. To ensure robustness, we replaced it with the more general Gaussian smearing method by setting ISMEAR = 0. In this process, we perform a self-consistent field (SCF) calculation. Based on the SCF results, we can make an initial estimation of the crystal’s direct and indirect band gaps, which were obtained by using the $Vasprun$ module in pymatgen to read the VASP output files. Since this is only a preliminary screening, we applied a loose threshold of 0.001 eV.

\subsubsection{Duplicate Structure Detection}

The duplication check encompasses structures from combined dataset\cite{heyu} of Materiae\cite{zhang2019catalogue} and TQC\cite{tqc}, and the newly generated crystals. To reduce computational cost, we first compare the chemical formulas of the unit cells. Only if they match do we proceed with $StructureMatcher$ module from pymatgen\cite{ong2013python} to determine whether two structures are similar, with parameters set to $ltol$ = 0.2, $stol$ = 0.3, and $angle\_tol$ = 5, which are also the default values. 

\subsubsection{Computation Runtime}
During high-throughput generation, the most time-consuming step is determining the topological properties of new materials using SymTopo, as this involves VASP calculations. Most materials can be processed within 10 CPU hours (on a 36-CPUs node, it typically takes about 10 to 20 minutes to classify one material using SymTopo). However, some materials are difficult to converge. Therefore, if the VASP calculation fails to converge within 36×2 CPU hours, we terminate the computation for that material.

During the crystal structure generation process with PODGen, we utilized an NVIDIA GeForce RTX 4090 GPU and approximately 20 CPU cores (with parallelization implemented via p\_tqdm in Python). The GPU was used for running the machine learning model, while the CPUs handled crystal graph construction, data format conversion, and related tasks.

The generation of several thousand new structures typically took about three days, which is relatively slow compared to standard generative models. This inefficiency is primarily attributed to two factors: (1) the use of MCMC-based sampling, which is inherently time-consuming; and (2) the low GPU utilization observed during sampling. We found that the majority of the computation time was spent on CPU-bound tasks such as crystal graph building and format conversions. This indicates significant room for efficiency improvements in our current implementation, particularly in optimizing the CPU-side workflow—an aspect we aim to address in the future.

Relaxing crystal structures using OpenLAM is highly efficient — with a single 36-core CPU node, we were able to complete the relaxation of approximately 900 structures within one hour.

\subsection{Additional Results}
\begin{figure}[h]
\includegraphics[width=0.45\textwidth]{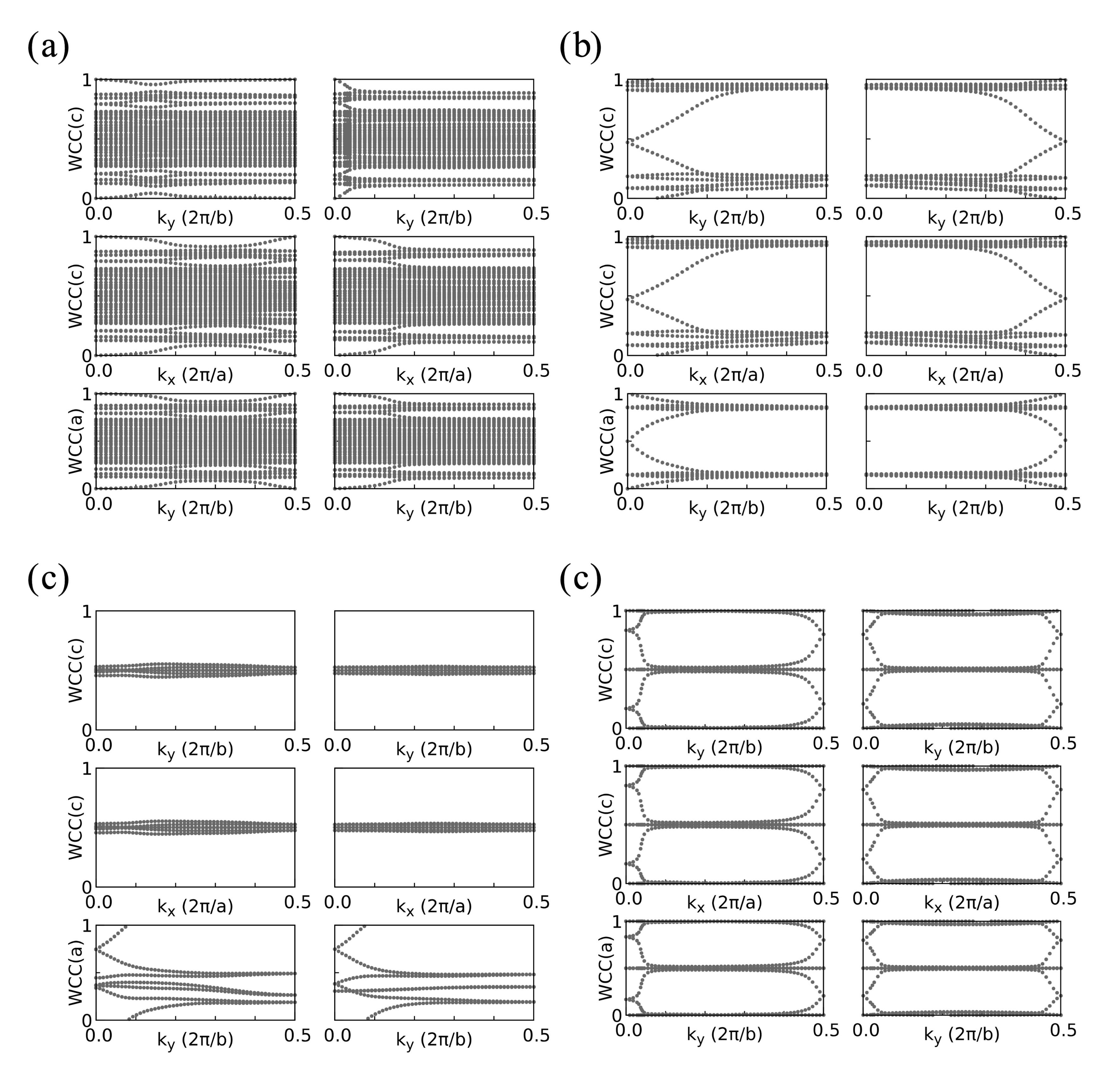} 
\caption{\label{fig:willson} $\mathbb{Z}_2$ invariants of (a) $\mathrm{BaMo_6As_2Se_6}$,(b) $\mathrm{Rb_2Sr_2Bi_2}$, and (c) $\mathrm{CsHgSb}$, obtained from the evolution of the Wannier charge centers (WCC). The result indicate $\mathbb{Z}_2$=(1;111) for $\mathrm{BaMo_6As_2Se_6}$, $\mathbb{Z}_2$=(0;111) for $\mathrm{Rb_2Sr_2Bi_2}$, $\mathbb{Z}_2$=(0;001) for $\mathrm{CsHgSb}$ and $\mathbb{Z}_2$=(0;000) for $\mathrm{Ca_2AgAs}$. Although $\mathrm{Ca_2AgAs}$ has a $\mathbb{Z}_2$ topological index of (0;000), its Wilson loop is not trivial, which can instead use $\mathbb{Z}_8$=4 \cite{song2018quantitative} to describe the nontrivial topological nature of the material.}
\end{figure}

\begin{figure*}[t]
\includegraphics[width=1\textwidth]{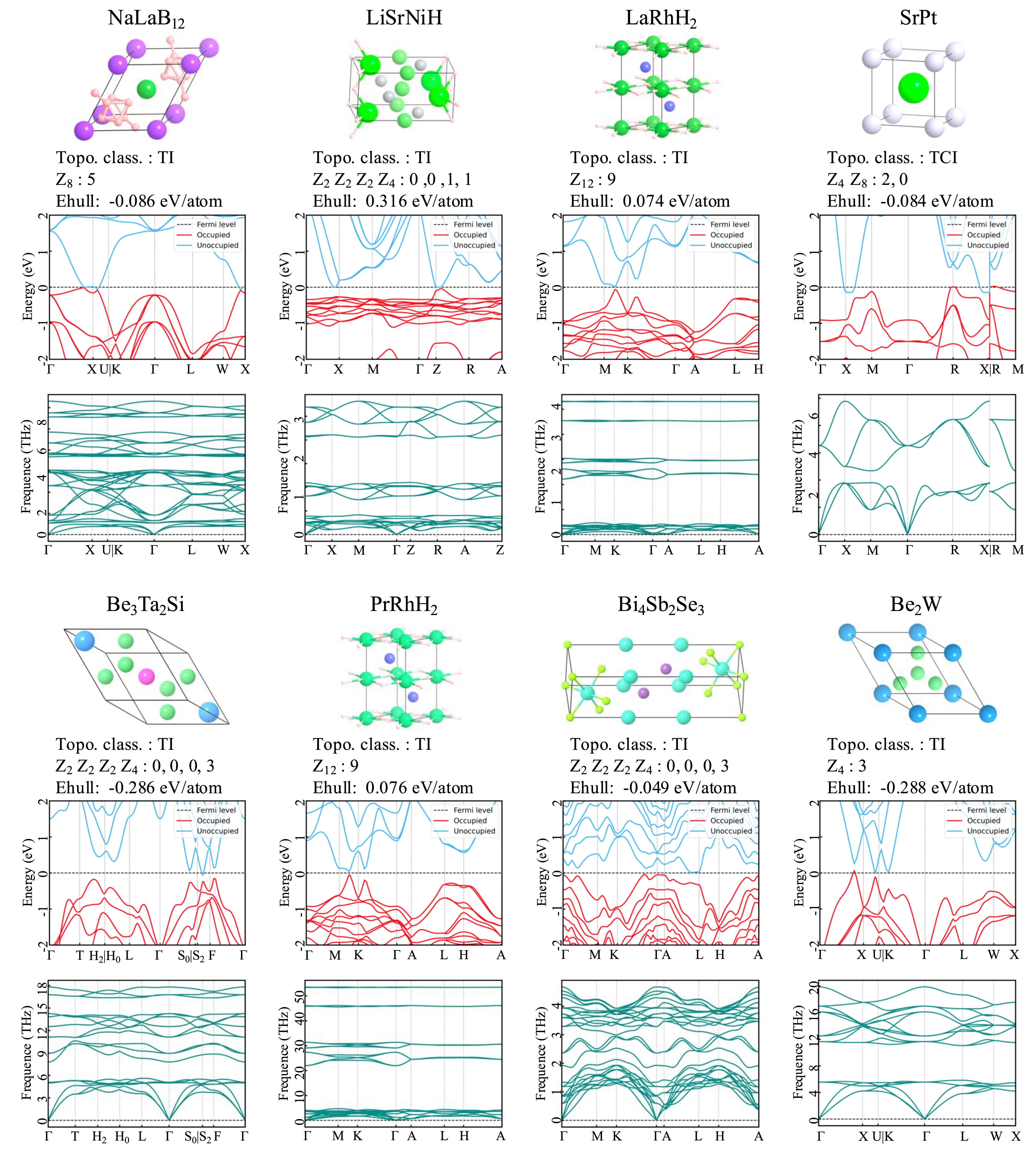}%
\caption{\label{fig:more_bs_phonon} The remaining eight crystal structures that retain their status as topological (crystalline) insulators with a direct band gap after further relaxation via DFT and exhibit phonon spectra without imaginary frequencies as determined by DFPT calculations. The topological classification and symmetry-based indicators for these structures are provided by SymTopo. Ehull represents the energy above hull, which are obtained from DFT calculations.}
\end{figure*}

\begin{figure*}[t]
\includegraphics[width=1\textwidth]{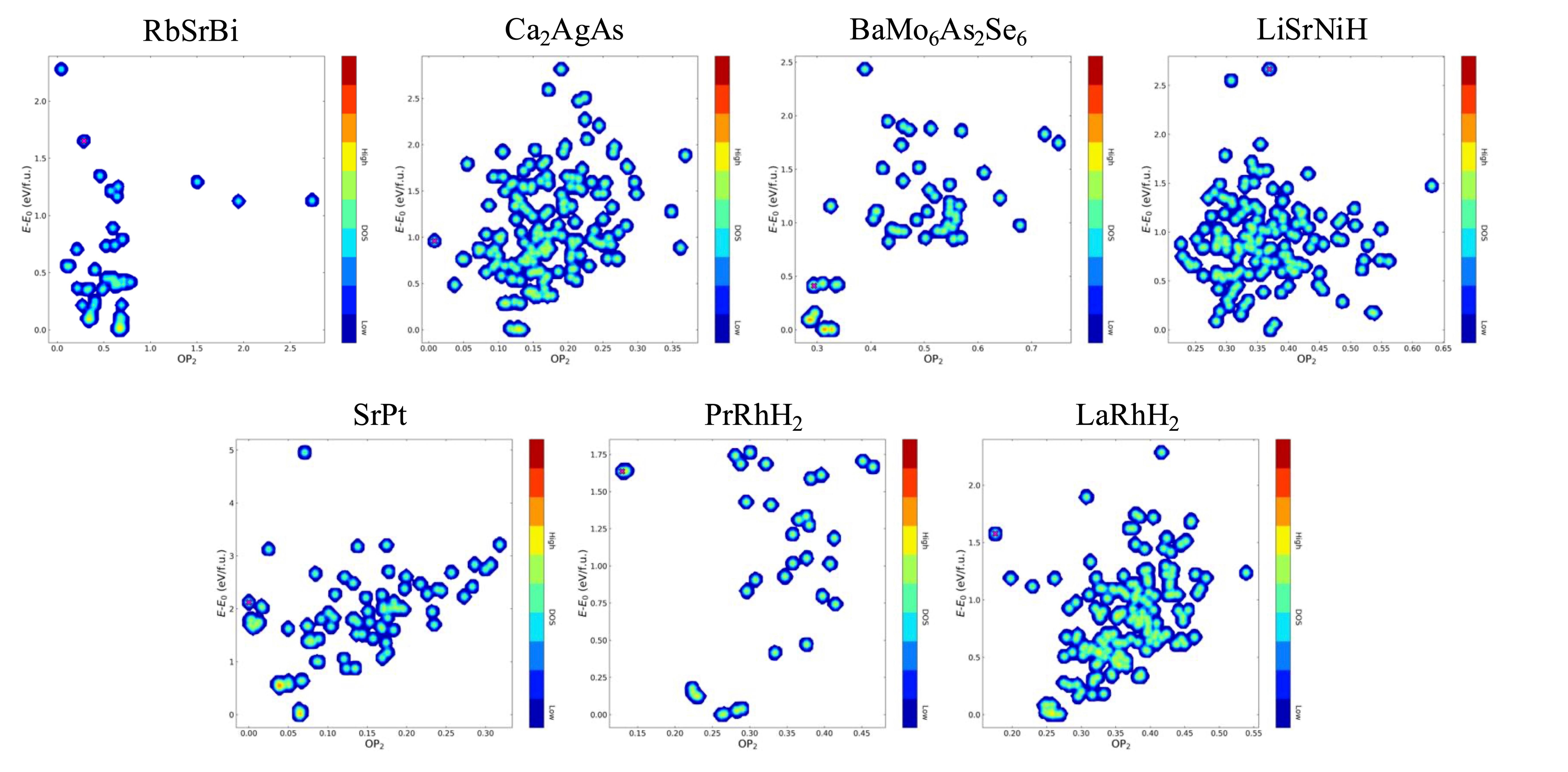}%
\caption{\label{fig:pessi} PES contour plots of seven newly generated materials obtained from SSW, where the vertical axis represents the relative energy and the horizontal axis denotes the structural descriptor. ‘DOS’ indicates the number of times each structure was found. The red crosses mark the positions of these five materials on their respective potential energy surfaces.}
\end{figure*}

Through further first-principles calculations, as outlined in FIG.~5, we identified 12 TI/TCI materials with a direct band gap and no imaginary frequencies in their phonon spectra. Eight of these materials are presented in FIG.~\ref{fig:more_bs_phonon}. And we use WannierTools to calculate the surface states and the Wilson loops. The detailed Wilson loop calculation results are shown in FIG.~\ref{fig:willson}. The potential energy surfaces of the remaining seven materials are shown in FIG.~\ref{fig:pessi}.

\begin{figure*}[t]
\includegraphics[width=1\textwidth]{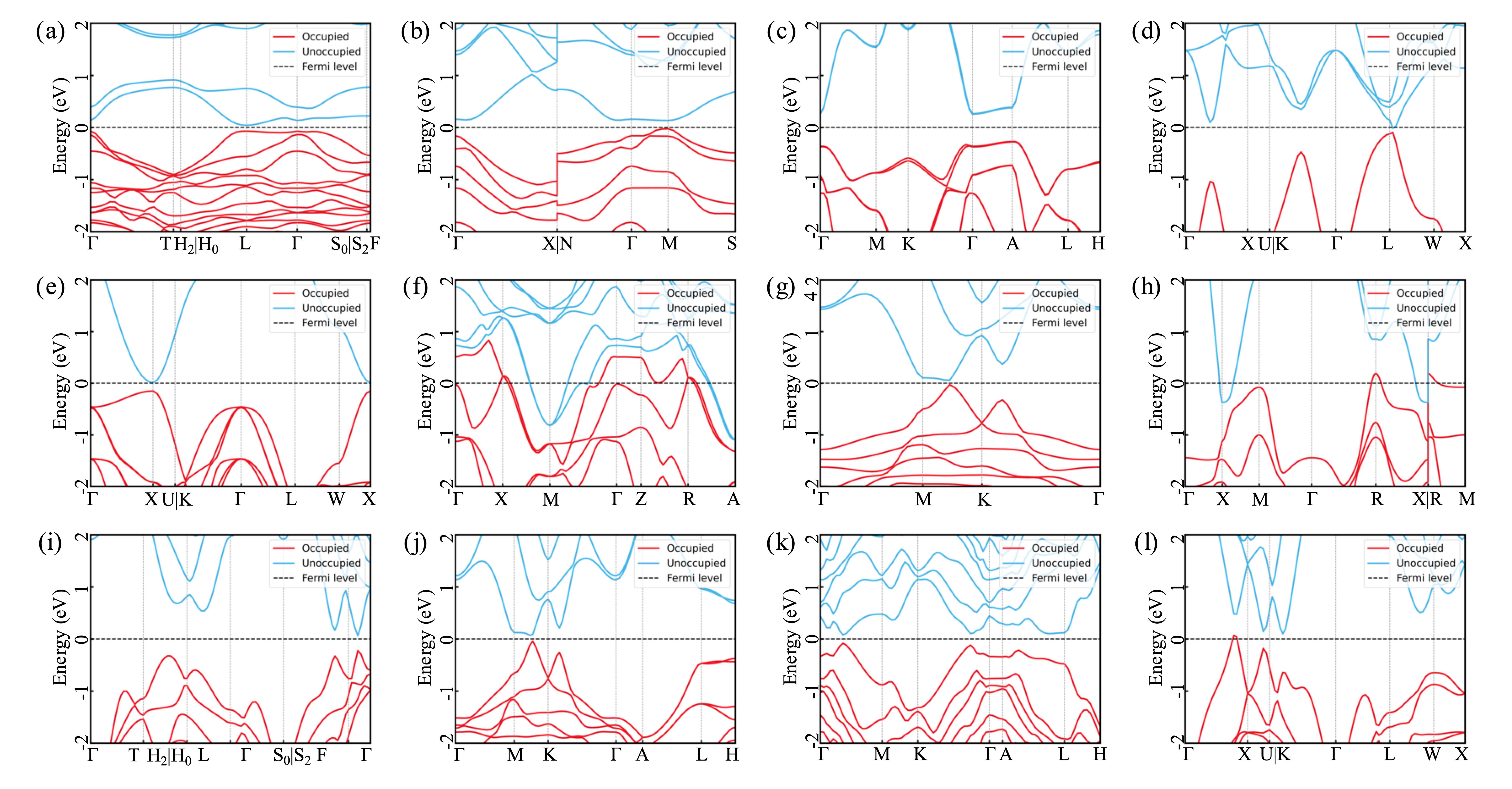}%
\caption{\label{fig:hse06} Band structures of 12 materials calculated using the HSE06 hybrid functional: (a)$\mathrm{BaMo_6As_2Se_6}$, (b)$\mathrm{RbSrBi}$, (c) $\mathrm{CsHgSb}$, (d)$\mathrm{Ca_2AgAs}$, (e)$\mathrm{NaLaB_{12}}$, (f)$\mathrm{LiSrNiH}$, (g)$\mathrm{LaRhH_{2}}$, (h)$\mathrm{SrPt}$, (i)$\mathrm{Be_{3}Ta_{2}Si}$, (j)$\mathrm{PrRhH_{2}}$, (k)$\mathrm{Bi_{4}Sb_{2}Se_{3}}$ and (l)$\mathrm{Be_{2}W}$.}
\end{figure*}

Based on the results obtained from PBE functional calculations, we recalculated the band structures of the twelve materials using the HSE06 hybrid functional\cite{hse06}, as shown in Fig.~\ref{fig:hse06}. The computational procedure involves performing a self-consistent field (SCF) calculation with the HSE06 functional, using the charge density (CHGCAR) obtained from the preceding PBE SCF calculation as the initial input, followed by a band structure calculation.

The PBE\cite{PBE} functional is known to underestimate the band gaps of semiconductors. In contrast, the HSE06 hybrid functional incorporates a more accurate treatment of exchange-correlation interactions, providing significantly improved band gap predictions. To ensure the robustness of our topological insulator characterization, we performed additional calculations using the HSE06 functional. The results show that the topological nature of most materials—such as $\mathrm{BaMo_6As_2Se_6}$, $\mathrm{RbSrBi}$, $\mathrm{CsHgSb}$, $\mathrm{Ca_2AgAs}$, $\mathrm{LaRhH_{2}}$, $\mathrm{SrPt}$, $\mathrm{Be_{3}Ta_{2}Si}$, $\mathrm{PrRhH_{2}}$, $\mathrm{Bi_{4}Sb_{2}Se_{3}}$ and $\mathrm{Be_{2}W}$—remains unchanged. However, two materials exhibit non-negligible differences. In the case of $\mathrm{NaLaB_{12}}$, the band inversion near the X point observed in the PBE results is no longer present in the HSE06 calculation, which may potentially alter its topological classification. In the case of $\mathrm{LiSrNiH}$, the previously flat band just below the Fermi level becomes substantially more dispersive under HSE06. This change may be attributed to the emergence of magnetic interactions involving the Ni atoms, which are not adequately captured in the non-magnetic PBE calculations.
